\documentclass[smallextended]{svjour3}

\smartqed  

\usepackage{graphicx}
\usepackage{amssymb, amsmath} 

\journalname{International Journal of Theoretical Physics}

\begin{document}

\title{Supersymmetric model of a Bose-Einstein condensate in a 
  $\mathcal{PT}$-symmetric double-delta trap}

\titlerunning{SUSY model of a BEC in a $\mathcal{PT}$-symmetric double-delta
  trap}

\author{Nikolas Abt \and Holger Cartarius \and G\"unter~Wunner}
\authorrunning{N. Abt, H. Cartarius, G. Wunner}

\institute{N. Abt \and H. Cartarius \and G. Wunner \at Universit\"at Stuttgart,
  Institut f\"ur Theoretische Physik 1, 70550 Stuttgart, Germany \and
  N. Abt \at \email{Nikolas.Abt@itp1.uni-stuttgart.de} \and
  H. Cartarius \at \email{Holger.Cartarius@itp1.uni-stuttgart.de} \and 
  G. Wunner \at \email{wunner@itp1.uni-stuttgart.de}}

\date{Received: date / Accepted: date}

\maketitle

\begin{abstract}
The most important properties of a Bose-Einstein condensate subject to
balanced gain and loss can be modelled by a Gross-Pitaevskii equation with an
external $\mathcal{PT}$-symmetric double-delta potential. We study its linear
variant with a supersymmetric extension. It is shown that both in the
$\mathcal{PT}$-symmetric as well as in the $\mathcal{PT}$-broken phase
arbitrary stationary states can be removed in a supersymmetric partner
potential without changing the energy eigenvalues of the other state. The
characteristic structure of the singular delta potential in the supersymmetry
formalism is discussed, and the applicability of the formalism to the
nonlinear Gross-Pitaevskii equation is analysed. In the latter case the
formalism could be used to remove $\mathcal{PT}$-broken states introducing
an instability to the stationary $\mathcal{PT}$-symmetric states.

\keywords{$\mathcal{PT}$ symmetry \and supersymmetry \and double-delta
  potential \and stationary states}
\PACS{03.75.Hh \and 11.30.Er \and 11.30.Pb \and 03.65.Ge}
\end{abstract}

\section{Introduction}
\label{sec:introduction}

Bose-Einstein condensates in a double-well setup, where in one well atoms
are extracted from the trap and in the other atoms are added coherently
to the condensed phase, have shown to be a good candidate for a realisation
of a $\mathcal{PT}$-symmetric quantum system, i.e.\ a system of which the
Hamiltonian commutes with the combined action of the parity and time-reversal
operators, $[\mathcal{PT},H]=0$ \cite{Klaiman08a,Dast13a,Graefe08b}. These
systems are of special interest since they allow for the existence of real
eigenvalues despite the presence of gain and loss of the probability amplitude,
which is described by the non-Hermitian contributions to the Hamiltonian
\cite{Bender98}. Real eigenvalues represent the situation of balanced gain and
loss such that a stationary probability distribution in the system of
interest exists.

While the existence of $\mathcal{PT}$-symmetric states has been shown
for Bose-Einstein condensates \cite{Dast13a,Graefe08b} there has to be
taken special care of the inter-atomic contact interaction. In the mean-field
limit of the Gross-Pitaevskii equation this interaction leads to a nonlinearity
in the Hamiltonian. Many aspects of $\mathcal{PT}$ symmetry in quantum systems,
e.g.\ the relation between $\mathcal{PT}$-symmetric eigenstates and real energy
eigenvalues, remain unchanged if the Gross-Pitaevskii nonlinearity $\propto 
|\psi|^2$ is added \cite{Dast13b}. However, the nonlinearity also leads to new
features. In linear quantum mechanics usually two energy eigenvalues approach
each other when the gain-loss effect is increased until they merge in an
exceptional point. For even stronger gain-loss contributions two complex
and complex conjugate eigenvalues belonging to $\mathcal{PT}$-broken wave
functions appear. In the nonlinear system these complex eigenvalues are not
born in the exceptional point, rather they branch off from one of the real
eigenvalues before it vanishes together with the second in the exceptional
point. In optical media, where a Kerr nonlinearity leads to a mathematical
description equivalent to that of the Gross-Pitaevskii equation, these effects
may be exploited for technical applications such as unidirectional wave guides
\cite{Ramezani10} or the propagation of solitons \cite{Musslimani08a,%
  Driben2011a,Bludov2013a}. However, the $\mathcal{PT}$-broken states also
introduce a dynamical instability to the stationary $\mathcal{PT}$-symmetric
state from which they branch off \cite{Haag14a,Loehle14a}.

The formalism of non-relativistic supersymmetry (SUSY) offers an elegant way of 
removing disturbing $\mathcal{PT}$-broken states without changing all other
states. In an experiment this possibility could be of great benefit. Initially
introduced in quantum field theories \cite{Gelfand1971a,Neveu1971a,Ramond1971a}
supersymmetry has also a large number of applications in non-relativistic
quantum mechanics \cite{Nicolai1976a,Witten1981a,Witten1982a}. A characteristic
property is the possibility to relate two quantum mechanical systems with
different potentials $V_1$ and $V_2$ by a supersymmetric transformation
such that they possess almost identical spectra. Apart from the ground state
of the system described by $V_1$ all eigenstates appear also in the system 
governed by $V_2$ with exactly the same energies but different wave functions.

Completely new perspectives are offered by supersymmetry in non-Her\-mi\-ti\-an
$\mathcal{PT}$-symmetric quantum systems. The relations of both symmetries
have been studied extensively \cite{Znojil2000a,Znojil2002a,Levai2002a}, where
much richer structures than in Hermitian systems and even purely real partner
potentials $V_2$ of complex $\mathcal{PT}$-symmetric potentials $V_1$ can be
found \cite{Bazeia2009a,Abhinav2010a,Levai2004a,Bagchi2000a}. In optics
the formalism has been used to study theoretically methods of designing the
refractive index of optical crystals such that they become unidirectionally
invisible \cite{Midya2014a} or to synthesise desired functionalities
\cite{Miri2013b}. Of particular interest for our purpose is the fact that
quantum systems described by a complex $\mathcal{PT}$-symmetric potential
$V_1$ can be related to a partner potential $V_2$ in which not only the ground
state but any arbitrary state can be removed. Miri et al.\ \cite{Miri2013a}
have shown that this property can be used to selectively remove unwanted modes
from a wave guide without hindering the propagation of desired waves. In this
article we want to extend this concept to matter waves.

A simple model that features all effects of a Bose-Einstein condensate in a
double-well with balanced gain and loss is the Gross-Pitaevskii equation of
the $\mathcal{PT}$-symmetric double-delta potential \cite{Cartarius12b,%
  Cartarius12c}. It is the main purpose of this paper to perform the first
step on the way to remove the $\mathcal{PT}$-broken states introducing the
dynamical instability. To do so, we apply the SUSY formalism to the
case of vanishing Gross-Pitaevskii nonlinearity. This potential has often
been used to gain deeper insight with analytically accessible energies or wave
functions \cite{Jakubsky2005a,Mehri-Dehnavi2010a,Jones2008a,%
  Mayteevarunyoo2008a,Rapedius08a,Fassari2012a}. We show that the SUSY scheme
can indeed be used to remove arbitrary $\mathcal{PT}$-symmetric and
$\mathcal{PT}$-broken eigenstates. The concept turns out to work well and
provides an infinite number of superpotentials for the removal of each
eigenstate. To understand the properties of supersymmetry in our system we
discuss in detail the characteristics of the singular delta potential, for
which so far mathematical investigations in the Hermitian case exist
\cite{Uchino2003a,Correa2008a,Fernandez2011a}. Furthermore, we comment on
possible extensions of the procedure to nonlinear systems and develop a method
of constructing a potential $V_2$ which for weak nonlinearities leads to good
approximate solutions.

The article is organised as follows. In Sect.\ \ref{sec:dd_ext} we
introduce the SUSY formalism and apply it to the $\mathcal{PT}$-symmetric
double-delta potential. Then we demonstrate how the procedure can be used to
remove an arbitrary eigenstate without influencing the remaining one in Sect.\
\ref{sec:removal}. We analyse in particular the case in which two states
coalesce at an exceptional point. The applicability of the formalism to
systems with a weak nonlinearity is discussed in Sect.\ \ref{sec:nonlin}.
Finally we summarise our results and give an outlook on possible extensions
of our approach to general nonlinearities in Sect.\ \ref{sec:conclusion}.

\section{Supersymmetric extension of the
  $\mathcal{PT}$-symmetric double-delta potential}
\label{sec:dd_ext}

In a first step we investigate how the SUSY formalism acts on the singular and
non-Hermitian $\mathcal{PT}$-symmetric double-delta potential. Since the
formalism was set up for linear quantum mechanics we take only into account
the linear parts of the Hamiltonian. Then we apply the standard scheme of
deriving the SUSY partner $\mathcal{H}_2$ (i.e.\ the Fermionic sector in SUSY
notation) of a given Hamiltonian $\mathcal{H}_1$ of the original system (Bosonic
sector). To do so, in terms of exact SUSY, the energy of $\mathcal{H}_1$ is
shifted such that the energy of the state we wish to remove is zero.
Thus, we consider the one-dimensional Schr\"odinger equation of the
$\mathcal{PT}$-symmetric double-delta potential in suitable units
\cite{Cartarius12b} and subtract the energy of the current eigenstate,
\begin{equation}
  \label{eq:onedgpe}
  \mathcal{H}_1 \phi_n^{(1)} = \left[ -\partial_x^2 
    + \nu \delta\left( x - \frac{a}{2} \right) + \nu^* \delta\left( x 
      + \frac{a}{2} \right) + \left( \kappa_n^{(1)} \right)^2 \right] 
  \phi_n^{(1)} = 0 \; ,
\end{equation}
where $ \phi_n^{(1)}$ is the eigenstate of $\mathcal{H}_1$ with the
corresponding eigenvalue $\mathcal{E}_n^{(1)}=-( \kappa_n^{(1)} )^2$. The complex
strength of the double-delta potential at $x=\pm a/2$ is denoted by $\nu
= -1 + \mathrm{i} \gamma$ and $\nu^* = -1 - \mathrm{i} \gamma$. The next step is
to factorise the Hamiltonian  by means of the creation and annihilation
operators $B^\pm$ to gain a link between the Bosonic and Fermionic sectors,
i.e. $\mathcal{H}_1$ and its supersymmetric partner Hamiltonian
$\mathcal{H}_2$, whose eigenstates and eigenenergies shall be calculated. To
this end we introduce
\begin{equation}
  \label{eq:creatorannihilator}
  B^\pm = \mathcal{W}(x) \mp \partial_x
\end{equation}
with the superpotential $\mathcal{W}(x)$. Using the canonical representation
both Hamiltonians can be combined in one SUSY Hamiltonian
\begin{align}
  \notag
  \mathcal{H}_S =&   \begin{pmatrix}
    \mathcal{H}_1 & 0 \\ 0 & \mathcal{H}_2 \end{pmatrix} = 
  \begin{pmatrix} B^+ B^- & 0 \\ 0 & B^- B^+ \end{pmatrix} \\
  =& \label{eq:susyhamiltonian}
  \begin{pmatrix}
    -\partial_x^2 + \mathcal{W}^2(x) -  \mathcal{W}'(x) & 0 \\
    0 & -\partial_x^2 +  \mathcal{W}^2(x) + \mathcal{W}'(x) 
  \end{pmatrix} \; .
\end{align}
Following the relations of quantum mechanical supersymmetry we identify 
\begin{equation}
  \label{eq:deq_superpot}
  V_1 = \mathcal{W}^2 - \mathcal{W}'
\end{equation}
with the present double delta-potential appearing in \eqref{eq:onedgpe}.
Consequently, we obtain a rule to calculate the superpotential,
\begin{equation}
  \label{eq:def_superpotential}
  \mathcal{W}(x) = - \frac{\partial_x \phi_n^{(1)}}{\phi_n^{(1)}} \; .
\end{equation}
Using the analytical solutions for $\phi_n^{(1)}$, c.f.\ Refs.\
\cite{Cartarius12b,Cartarius12c}, the superpotential is given by
\begin{equation}
  \label{eq:superpotential}
  \mathcal{W} = \begin{cases}
    - \kappa_n^{(1)}  \quad & \text{for } x<-\frac{a}{2}\; ,\\
    -\kappa_n^{(1)} \frac{1 + \left(1 + 2 \kappa_n^{(1)}/ \nu \right)
      \exp \left(-\kappa_n^{(1)}(2x -a) \right)}{1-\left( 1+ 2 \kappa_n^{(1)}/ 
        \nu \right) \exp \left(-\kappa_n^{(1)}(2x -a) \right)}  \quad & 
    \text{for } -\frac{a}{2} < x < \frac{a}{2} \; ,\\
    \kappa_n^{(1)}	 \quad & \text{for } x>\frac{a}{2} \; .
  \end{cases}
\end{equation}
Equation \eqref{eq:superpotential} predicts a jump in the superpotential and
therefore a divergence in its first derivative at the positions of the delta
functions. Since we use $\mathcal{W}$ and $\mathcal{W}'$ to generate the
supersymmetric partner potential $V_2$ and calculate the eigenfunctions of
$\mathcal{H}_2$ it is necessary to know the impact on the solutions of
$\mathcal{H}_2$. Hence, we have to understand the appearance of the
delta-singularity in detail.

Starting with the stationary Schr\"odinger equation of the Fermionic sector
\begin{equation}
  \label{eq:sesystemtwo}
  \left( \partial_x^2 - \left[ \mathcal{W}^2 + \mathcal{W}' \right] \right)
  \phi_n^{(2)} = \mathcal{E}_n^{(2)} \phi_n^{(2)} 
\end{equation}
we find for the jump in the first derivative of the wave function by
integrating over a small neighbourhood around the position of the delta
function at $x=+a/2$
\begin{equation}
  \label{eq:jumpwavefunctiontwo:1}
  \Delta  \partial_x \phi_n^{(2)} = \lim_{\varepsilon \rightarrow 0} \int_{\frac{a}{2}
    - \varepsilon}^{\frac{a}{2} + \varepsilon} \text{d}x \, \mathcal{W}'
  \phi_n^{(2)} \; .
\end{equation}
These are the only terms which remain in the integral. All other terms
of \eqref{eq:sesystemtwo} are continuous and vanish in the limit $\varepsilon
\to 0$. Furthermore we can calculate the jump in the derivative of the
superpotential from its definition in \eqref{eq:def_superpotential},
\begin{align}
  \notag
  \Delta  \mathcal{W} \left( \frac{a}{2} \right) &= \lim_{\varepsilon \rightarrow 0}
  \int_{\frac{a}{2}-\varepsilon}^{\frac{a}{2}+\varepsilon} \text{d} x \, \partial_x
  \left[- \frac{\partial_x \phi_n^{(1)}}{\phi_n^{(1)}} \right]\\
  \notag
  &= - \lim_{\varepsilon \rightarrow 0} \left[ \frac{\partial_x \phi_n^{(1)}
      \left( \frac{a}{2}+\varepsilon \right)}{\phi_n^{(1)}\left( 
        \frac{a}{2}+\varepsilon \right) } - \frac{\partial_x \phi_n^{(1)}\left(
        \frac{a}{2}-\varepsilon \right)}{\phi_n^{(1)}\left( \frac{a}{2}
        -\varepsilon \right) } \right]\\
  &= - \frac{1}{\phi_n^{(1)}\left( \frac{a}{2} \right)}\Delta \partial_x
  \phi_n^{(1)}\left( \frac{a}{2} \right) = - \nu \; ,
  \label{eq:jumpsuperpotential}
\end{align}
where we used the result for the jump in the derivative of the wave function
in a double-delta potential. This result can be used to formally substitute
the occurrence of $\mathcal{W}'$ with delta functions. The situation is
symmetric, therefore we get a similar result for the integration at $x=-a/2$,
namely $\Delta \mathcal{W} (-a/2) = -\nu^*$. By using
\begin{equation}
  \label{eq:substitutionwprime}
  \mathcal{W}' \rightarrow - \nu \delta \left( x - \frac{a}{2} \right) - \nu^* \delta \left( x + \frac{a}{2} \right) \; ,
\end{equation}
equation \eqref{eq:jumpwavefunctiontwo:1} assumes the shape
\begin{equation}
  \label{eq:jumpwavefunctiontwo:2}
  \Delta \partial_x \phi_n^{(2)} \left( \frac{a}{2} \right) = - \nu
  \phi_n^{(2)} \left( \frac{a}{2} \right) \; , \quad \Delta \partial_x
  \phi_n^{(2)}  \left( - \frac{a}{2} \right) = - \nu^* \phi_n^{(2)} \left(
    - \frac{a}{2} \right) \; .
\end{equation}
Given all the expressions above, we are able to provide an analytical
solution for the partner potential $V_2$, viz.\
\begin{multline}
  \label{eq:potentialv2}	
  V_2(x) = - \nu \delta \left( x - \frac{a}{2} \right) - \nu^* \delta
  \left( x + \frac{a}{2} \right)\\
   \hspace*{-30pt} +	\begin{cases}
    \left( \kappa_{n}^{(1)} \right)^2 \quad &\text{for } x<-\frac{a}{2}\, ,\\
    \left( \kappa_{n}^{(1)} \right)^2  \left[ \frac{1 + \left(1 + 2
          \kappa_n^{(1)}/ \nu \right) \exp \left(-\kappa_n^{(1)}(2x -a)
        \right)}{1-\left( 1+ 2 \kappa_n^{(1)}/ \nu \right) \exp \left(
          -\kappa_n^{(1)}(2x -a) \right)} \right]^2 \quad &\text{for } 
    -\frac{a}{2}<x<\frac{a}{2}\, ,\\
    \left( \kappa_{n}^{(1)} \right)^2 \quad &\text{for } x>\frac{a}{2}\, .
  \end{cases}			
\end{multline}

\section{Removal of arbitrary eigenstates in the linear
  system}
\label{sec:removal}

In the previous section we showed that the superpotential and the
supersymmetric partner $V_2$ of the $\mathcal{PT}$-symmetric double-delta
potential $V_1$ can be found in an analytic way if the corresponding eigenvalue 
$\kappa_{n}^{(1)}$ of the system $V_1$ is known. This eigenvalue has to be
determined numerically. To do so, we solve the ordinary differential equation
\eqref{eq:onedgpe} by integrating the wave function outward from $x=0$ in
positive and negative directions. Since its global phase is arbitrary
the three initial values $\mathrm{Re}\, \phi_n^{(1)}(0)$, $\mathrm{Re}\,
\phi_n^{(1)\prime}(0)$, and $\mathrm{Im} \, \phi_n^{(1)\prime}(0)$ together with
an estimate for the complex number $\kappa_n^{(1)}$ are determined in a
five-dimensional root search such that physically relevant wave functions are
obtained \cite{Cartarius12b}. They have to be normalised (one real condition)
and their real and imaginary parts have to vanish in the limits $x \to \pm
\infty$ (four conditions). Numerical solutions of the system $V_2$ are then
found by applying the same technique to the Schr\"odinger equation
\begin{equation}
  \mathcal{H}_2 \phi_n^{(2)} = \left[ -\partial_x^2 + V_2(x) \right] 
  \phi_n^{(2)} = \kappa_n^{(2)}  \phi_n^{(2)}
\end{equation}
with $V_2(x)$ as defined in \eqref{eq:potentialv2}.

\begin{figure}
  \centering
  \includegraphics[width=0.6\textwidth]{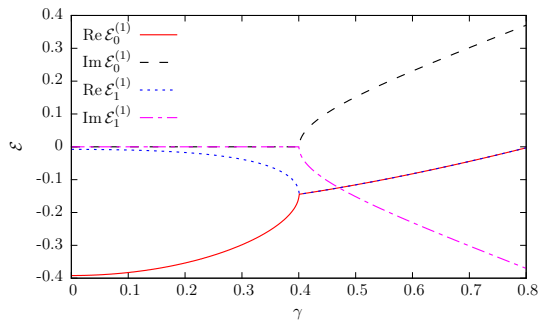}
  \caption{\label{fig:eigenvalues_original}Energy eigenvalues $\mathcal{E}_n
    = -(\kappa_n^{(1)} )^2$ of the ground state ($n=0$) and the
    excited state ($n=1$). Below a critical value $\gamma_\mathrm{crit}$ of
    the non-Hermiticity parameter $\gamma$ both energies are real, at
    $\gamma_\mathrm{crit}$ they merge in an exceptional point and above this
    value of $\gamma$ both energies are complex and complex conjugate.}
\end{figure}

\begin{figure}
  \includegraphics[width=\textwidth]{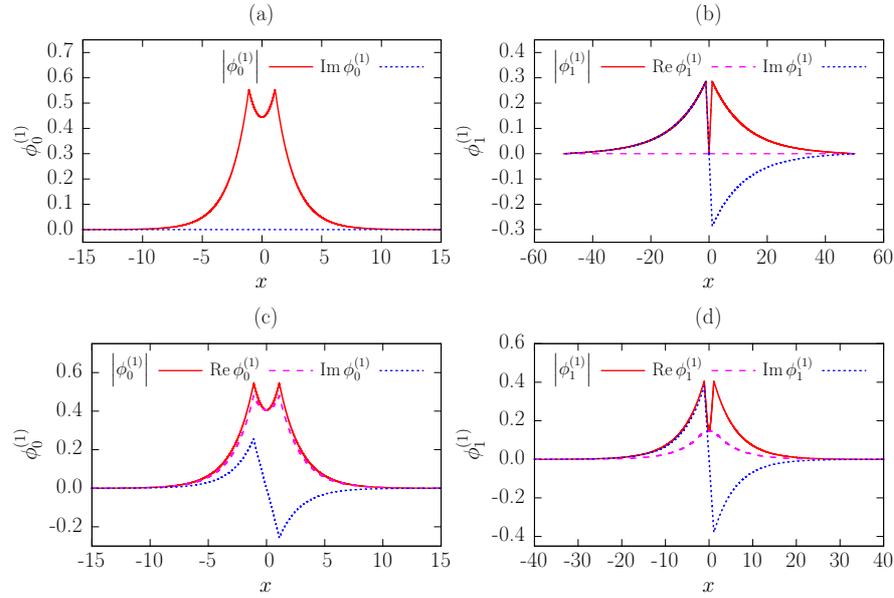}
  \caption{\label{fig:waves_original}Wave functions of the ground state 
    $\phi_0^{(1)}$ for $\gamma = 0$ (a), the excited state for $\gamma = 0$ (b),
    the ground state for $\gamma = 0.3$ (c), and the excited state for
    $\gamma = 0.3$ (d). Shown are the real and imaginary parts as well as
    the moduli. In (a) the real part coincides with the modulus and is not
    shown.}
\end{figure}

The original system $\mathcal{H}_1$ shows a typical spectrum of a
$\mathcal{PT}$-symmetric quantum system with two eigenstates, which is shown
in Fig.\ \ref{fig:eigenvalues_original}.
For a purely Hermitian potential, i.e.\ $\gamma = 0$, we obtain two real
energy eigenvalues, which remain real for increasing $\gamma$ until a critical
value $\gamma_\mathrm{crit} \approx 0.4005$ is reached. Their wave functions
are $\mathcal{PT}$ symmetric. At the critical value both eigenstates merge in
an exceptional point, i.e.\ their energies and wave functions coalesce. Above
$\gamma_\mathrm{crit}$ the two energies are complex and complex conjugate. The
corresponding wave functions are $\mathcal{PT}$ broken. Throughout this article
the wave function $\phi_0^{(1)}$ labels the ground state below
$\gamma_\mathrm{crit}$ and the state with positive imaginary part of the energy
above $\gamma_\mathrm{crit}$. The excited state or the state with negative
imaginary part of the complex energy are denoted by $\phi_1^{(1)}$. The wave
functions $\phi_0^{(1)}$ and $\phi_1^{(1)}$ are drawn in Fig.\ 
\ref{fig:waves_original}.

\subsection{Removal of $\mathcal{PT}$-symmetric states}

\begin{figure}
  \centering
  \includegraphics[width=0.6\textwidth]{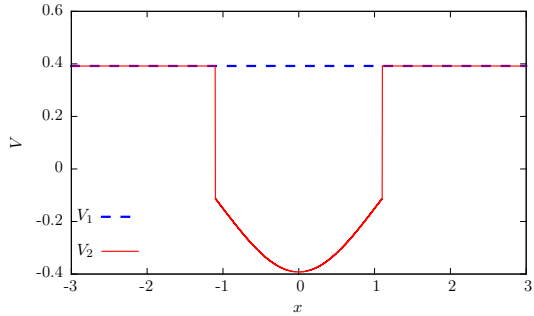}
  \caption{\label{fig:ground_state_V2}Potentials $V_1$ and $V_2$ for
    the SUSY formalism applied to the ground state $\phi_0^{(1)}$ in
    the case $\gamma = 0$.}
\end{figure}

First we concentrate on the spectrum for values of the non-Hermiticity
parameter below the critical value $\gamma_\mathrm{crit}$, which is most
illustrative since the eigenvalues $\kappa_n^{(1)}$ remain real and the
superpotential $\mathcal{W}$ and the potential $V_2$ of the Fermionic sector
can be chosen to preserve $\mathcal{PT}$ symmetry. If we use the ground state
$\phi_0^{(1)}$ for the construction of the supersymmetric partner we obtain the
potentials $V_1$ and $V_2$, which are shown in Fig.\ \ref{fig:ground_state_V2}
without the singular delta contributions. The potential $V_1$ is shifted by
the value of the original ground state's energy $\mathcal{E}_0^{(1)} = - (
\kappa_0^{(1)} )^2$ to be in the case of exact SUSY. More interesting is the
shape of $V_2$. It contains an additional symmetric potential well between
the \emph{repulsive} [cf.\ the analytic form in \eqref{eq:potentialv2}] delta
functions. This leads to completely different forms of the wave functions.

\begin{figure}
  \includegraphics[width=\textwidth]{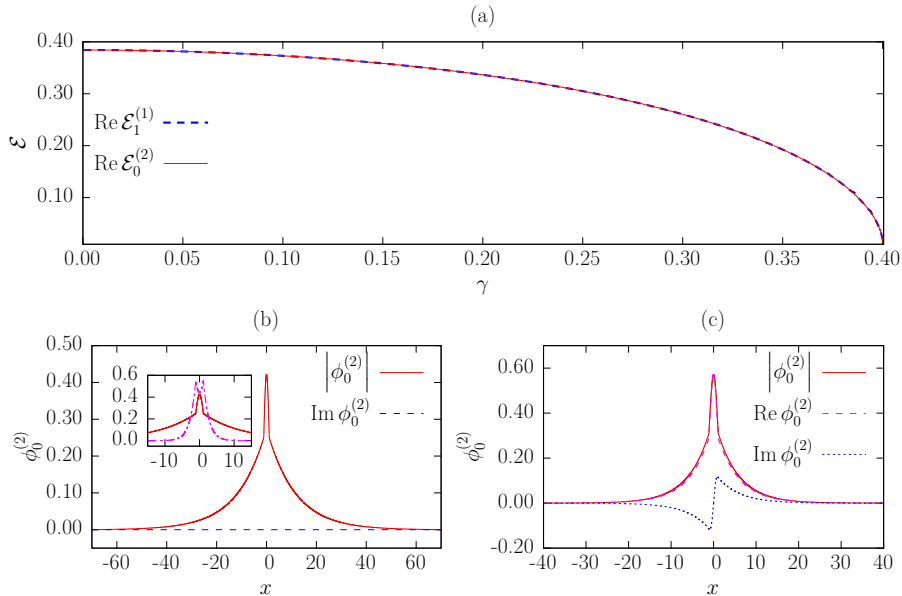}
  \caption{\label{fig:removed_groundstate}(a) Energy spectrum 
    $\mathcal{E}_0^{(2)}$ of the only existing state $\phi_0^{(2)}$ in the
    system described by the potential $V_2$ for removed ground state. For
    comparison the energy $\mathcal{E}_1^{(1)}$ of the original system is also
    shown. (b) Modulus (solid line) and imaginary part (dashed line) of the
    wave function $\phi_0^{(2)}$ in the case $\gamma = 0$. The real part
    coincides with the modulus and is not shown. The inset provides a direct
    comparison of $\phi_0^{(2)}$ (solid line) with the ground state
    $\phi_0^{(1)}$ of the original system (dashed-dotted line). (c)
    Modulus (solid line), real (dashed line) and imaginary (dotted line) parts
    of the wave function $\phi_0^{(2)}$ for $\gamma = 0.3$. In (b) and (c) the
    deviations from the shapes of the wave functions with the same energy
    eigenvalues in Figs.\ \ref{fig:waves_original}(b) and (d) are clearly
    visible.}
\end{figure}

Due to the SUSY formalism the ground state is removed in the system described
by $V_2$. The only existing state is the former excited state, which is the
ground state of the new system, and hence is labelled $\phi_0^{(2)}$.
Its energy must be positive since it must be above that of $\phi_0^{(1)}$,
which was set to zero in \eqref{eq:onedgpe}. Additionally we expect 
$\mathcal{E}_0^{(2)} < | \mathcal{E}_0^{(1)} |$ because the state
has to be bound. This is exactly what is found in our numerical solution.
Figure \ref{fig:removed_groundstate}(a)
shows the energy $\mathcal{E}_0^{(2)}$. It is always real, positive, and
slightly below $| \mathcal{E}_0^{(1)} |$. As $\gamma$ approaches
$\gamma_\mathrm{crit}$ and both states of the original system begin to merge we
observe the expected behaviour $\mathcal{E}_0^{(2)} \to 0$.

The numerical solution for the wave function at $\gamma = 0$ is depicted in
Fig.\ \ref{fig:removed_groundstate}(b). The differences to the wave functions
with the same energy eigenvalue of the original system in Figs.\
\ref{fig:waves_original}(b) and (d) are obvious. In particular, the state
is a true symmetric ground state of the partner system $V_2$, whereas in the
original system the state with exactly the same energy eigenvalue was the
antisymmetric excited state. Due to the attractive well around the origin the
wave function $\phi_0^{(2)}$ has its maximum at $x=0$ and decays as $x$
increases. The binding energy of this state $|\mathcal{E}_0^{(1)} | - 
\mathcal{E}_0^{(2)} \approx 0.0077$ is much lower than that of the
ground state of the original system, which has the value $| \mathcal{E}_0^{(1)}|
\approx 0.3920$. Consequently $\phi_0^{(2)}$ is considerably less localised
than the ground state $\phi_0^{(1)}$, which can be seen in the direct comparison
in the inset of Fig.\ \ref{fig:removed_groundstate}(b).

As is known from the original $\mathcal{PT}$-symmetric potential the 
antisymmetric imaginary part of the ground state's wave function grows in
strength for increasing $\gamma$. The same behaviour is observed for the ground
state $\phi_0^{(2)}$ of the Fermionic sector as can be seen in
\ref{fig:removed_groundstate}(c). Thus, we observe up to the critical value
$\gamma_\mathrm{crit}$ the typical behaviour of a $\mathcal{PT}$-symmetric
quantum system with the peculiarity that our potential exhibits only a single
bound state.

\begin{figure}
  \includegraphics[width=\textwidth]{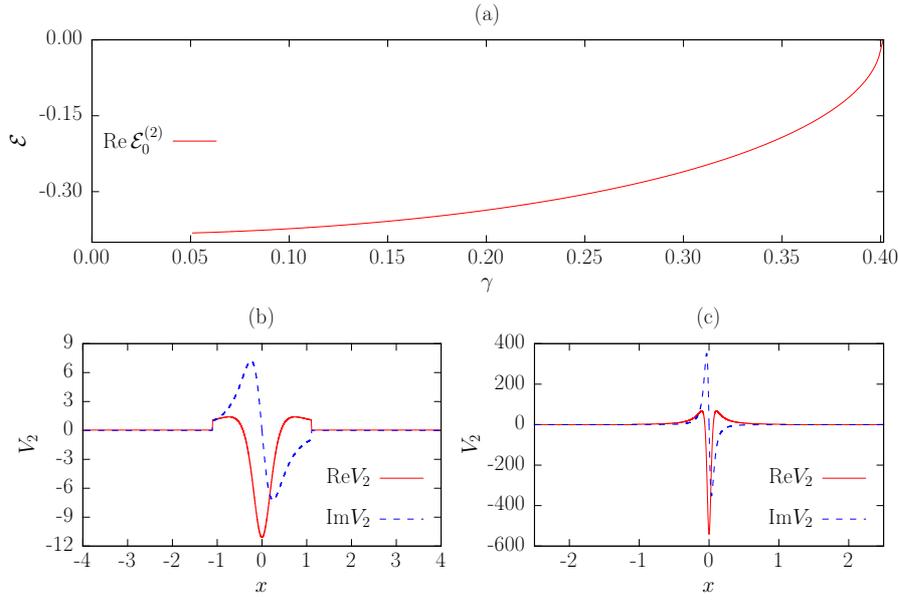}
  \caption{\label{fig:removed_exstate}(a) Energy spectrum $\mathcal{E}_0^{(2)}$
    of the only existing state $\phi_0^{(2)}$ in the system described by the
    potential $V_2$ for removed excited state up to $\gamma =
    \gamma_\mathrm{crit}$. (b) Real (solid line) and imaginary (dashed line)
    parts of the potential $V_2$ for $\gamma = 0.3$. It can be seen that the
    potential is $\mathcal{PT}$ symmetric. (c) Potential $V_2$ for $\gamma
    = 0.05$, where it already begins to diverge.}
\end{figure}

As was mentioned above the special feature of SUSY in $\mathcal{PT}$-symmetric
quantum systems is the possibility to remove an arbitrary state from the
spectrum of the partner potential $V_2$ provided that its wave function
does not have a node. We demonstrate this in our model by removing the energy
of the excited state $\phi_1^{(1)}$ for $\gamma \neq 0$. This is achieved
with exactly the same procedure as for the ground state with the sole difference
that now the eigenvalue $\kappa_1^{(1)}$ is used in the construction of the
potential $V_2$ according to \eqref{eq:def_superpotential} and 
\eqref{eq:sesystemtwo}. The spectrum is shown in Fig.\
\ref{fig:removed_exstate}(a).
The potential $V_2$ for $\gamma = 0.3$ can be seen in Fig. 
\ref{fig:removed_exstate}(b). This example demonstrates the $\mathcal{PT}$
symmetry of the potential. The partner system can be calculated numerically
only up to a minimal value of $\gamma$ because for $\gamma \to 0$ the
wave function $\phi_1^{(1)}$ approaches more and more the exact shape of the
antisymmetric ground state with its node at the origin. This node is reflected
in the potential $V_2$, which diverges in this limit at $x = 0$. Already for
$\gamma = 0.05$ the real part assumes a minimum value of $V_2(0) \approx -540$,
which can be observed in Fig.\ \ref{fig:removed_exstate}(c). This divergence
is not surprising since the SUSY formalism is expected to fail for the removal
of the excited state in a Hermitian quantum system.

\subsection{Removal of $\mathcal{PT}$-broken states}

\begin{figure}
  \centering
  \includegraphics[width=0.6\textwidth]{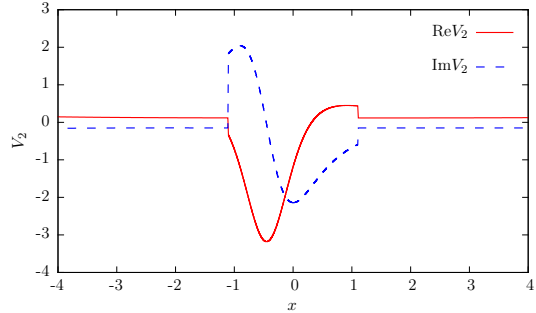}
  \caption{\label{fig:V2_broken}Real (solid line) and imaginary (dashed line)
    parts of the potential $V_2$ for $\gamma = 0.5$ and removed state
    $\phi_0^{(1)}$ with positive imaginary part of the energy. The potential
    is no longer $\mathcal{PT}$ symmetric.}
\end{figure}

The construction of a Fermionic sector for our model system in the
$\mathcal{PT}$-broken phase is no difficulty. The partner potential $V_2$
from \eqref{eq:potentialv2} remains valid in this case. Only the eigenvalue
$\kappa_n^{(1)}$, which appears in the equation, is now complex. An immediate 
consequence is the loss of $\mathcal{PT}$ symmetry of the partner potential,
which is illustrated in Fig. \ref{fig:V2_broken},
in which the potential $V_2$ is drawn for the removal of the state $\phi_0^{(1)}$
at $\gamma = 0.5$, i.e.\ beyond the exceptional point.

\begin{figure}
  \centering
  \includegraphics[width=0.6\textwidth]{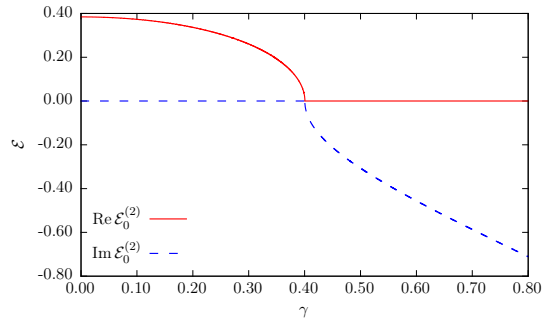}
  \caption{\label{fig:energies_removed_phi_1}Real (solid line) and imaginary
    (dashed line) parts of the energy $\mathcal{E}_0^{(2)}$ for removed
    state $\phi_0^{(1)}$ on both sides of the exceptional point.}
\end{figure}

The spectrum for the removal of $\phi_0^{(1)}$ on both sides of the exceptional
point can be seen in Fig.\ \ref{fig:energies_removed_phi_1}.
Also in the $\mathcal{PT}$-broken phase only one state remains. The purely
imaginary energy with $\mathrm{Im} \, \mathcal{E}_0^{(2)} < 0$ can be understood
if one remembers that due to exact SUSY the energy of the original system has
been shifted such that the energy of the removed state is set to zero. Thus we
expect to observe $\mathcal{E}_0^{(2)} = \mathcal{E}_1^{(1)} - \mathcal{E}_0^{(1)}$
in Fig.\ \ref{fig:energies_removed_phi_1}. This is exactly what is found. In the
original system the energies of both eigenstates of the double-delta potential
are complex conjugate and we can calculate
\begin{equation}
  \mathcal{E}_0^{(2)} = \mathcal{E}_1^{(1)} - \mathcal{E}_0^{(1)} = 
  2 \mathrm{i} \, \mathrm{Im} \, \mathcal{E}_1^{(1)}
\end{equation}
with $\mathrm{Im} \, \mathcal{E}_1^{(1)} = - \mathrm{Im} \, \mathcal{E}_0^{(1)}
< 0$.

\subsection{Behaviour at the exceptional point}

\begin{figure}
  \centering
  \includegraphics[width=0.6\textwidth]{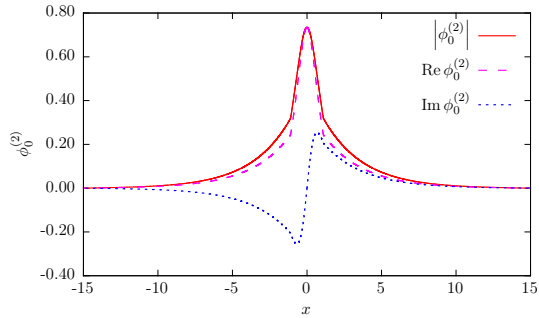}
  \caption{\label{fig:wave_gammacrit}Modulus (solid line) as well as
    real (dashed line) and imaginary (dotted line) parts of the wave
    function $\phi_0^{(2)}$ at the exceptional point at $\gamma_\mathrm{crit}
    \approx 0.4005$.}
\end{figure}

In the spectrum in Fig.\ \ref{fig:energies_removed_phi_1} we also observe that
$\mathcal{E}_0^{(2)} = 0$ at $\gamma_\mathrm{crit}$. This is not surprising since
at the exceptional point both energies of the original system coincide.
However, there remains one question. The SUSY formalism -- as it is introduced
in Hermitian quantum mechanics -- can be used to remove the ground state of
$\mathcal{H}_1$. Since exactly at the exceptional point also the original
$\mathcal{PT}$-symmetric double-delta potential $V_1$ has only \emph{one}
linearly independent state there should exist \emph{no} wave function at
$\gamma = \gamma_\mathrm{crit}$. However, this is not the case. Numerically we
find a solution at the critical value of $\gamma$, which is
$\mathcal{PT}$ symmetric. Its wave function is shown in Fig.\
\ref{fig:wave_gammacrit}.

There are two possible interpretations of this fact. Firstly, we may assume
that the supersymmetry formalism fails in the construction of a true Fermionic
sector if the potential $V_1$ exhibits coalescing eigenstates. Secondly, we
may interpret the coalescence at the exceptional point as two individual wave
functions which are just equal. Then one may argue that one of these wave
functions vanishes, whereas the second survives in the Fermionic sector and
supersymmetry is broken. For this interpretation one has to circumvent the
difficulty that for broken supersymmetry no state with $\mathcal{E}_n^{(2)} = 0$
may exist, which can be achieved by giving up the energy shift required for
exact SUSY. However, both possibilities are only interpretations which as a
matter of principle cannot be distinguished. The important fact is that,
independently from the state $\phi_n^{(1)}$ which is removed, the potential
$V_2$ \emph{always} exhibits \emph{one} eigenstate, even at the exceptional
point.

\subsection{Infinitely many superpotentials and real
  eigenvalues in non-$\mathcal{PT}$-symmetric potentials}
\label{sec:integration_constant}

In Sect.\ \ref{sec:dd_ext} we introduced the superpotential $\mathcal{W}(x)$
in the standard form shown in \eqref{eq:def_superpotential}. However, this is
only one possible solution of the differential equation \eqref{eq:deq_superpot}.
Its solutions possess an arbitrary integration constant. We mentioned above
that in the $\mathcal{PT}$-symmetric phase of the Bosonic sector
$\mathcal{H}_1$ the potential $V_2$ of the Fermionic sector preserves 
$\mathcal{PT}$ symmetry. This is the case because the $\mathcal{PT}$-symmetric
wave function $\phi_n^{(1)}$ of $\mathcal{H}_1$ is used in the construction
of $\mathcal{W}(x)$ in \eqref{eq:def_superpotential}, which chooses the
integration constant appropriately. In general, $V_2$ will not be
$\mathcal{PT}$ symmetric. Since every $\mathcal{H}_2$ must be isospectral with
that chosen according to the standard form \eqref{eq:def_superpotential} it
will possess only one eigenstate with \emph{real} energy in spite of the fact
that it is neither Hermitian nor $\mathcal{PT}$ symmetric. Thus, the
exploitation of the freedom of the integration constant offers one possibility
to construct Hamiltonians which do not posses a special symmetry but exhibit
real eigenvalues and even purely real spectra. This finding is equivalent to
that discussed in \cite{Miri2013a}.

Integration of the differential equation \eqref{eq:deq_superpot} with the
potential
\begin{equation}
  V_1(x) = -\mathcal{E}_1 = \left ( \kappa_n^{(1)} \right )^2
\end{equation}
outside and inside the delta functions leads to the superpotential
\begin{equation}
  \mathcal{W}(x) = -\kappa_n^{(1)} \tanh \left ( \kappa_n^{(1)} ( x -\xi) 
  \right ) \; ,
\end{equation}
where $\xi$ is a complex integration constant. One of these solutions is always
found if the differential equation \eqref{eq:deq_superpot} is solved
numerically. Note that the form presented in Eqs.\ \eqref{eq:superpotential}
and \eqref{eq:potentialv2} corresponds to $\xi \to \infty$ and $\xi \to 
-\infty$ in the intervals $x > a/2$ and $x < -a/2$, respectively.

\begin{figure}
  \includegraphics[width=\textwidth]{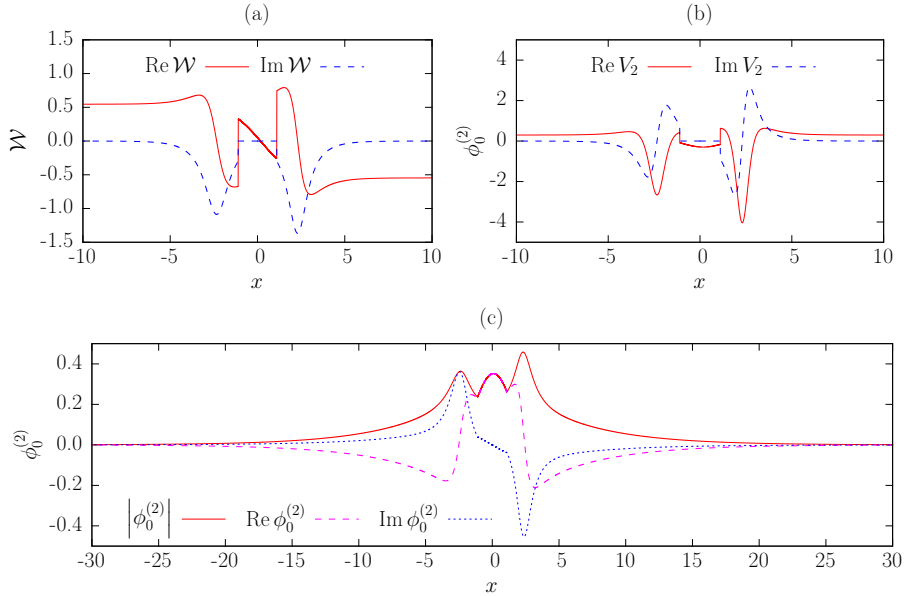}
  \caption{\label{fig:non-pt}(a) Real (solid line) and imaginary
    (dashed line) parts of the superpotential for the removal of the
    ground state and $\gamma = 0.3$. The values of the integration constant
    $\xi$ are $\xi = 2.30+2.18 \mathrm{i}$ and $\xi = -2.34+2.02\mathrm{i}$
    in the intervals $x > a/2$ and $x < -a/2$, respectively. (b) Potential
    $V_2$ for the same case. (c) Wave functions of the eigenstate $\phi_0^{(2)}$
    obtained with this potential.}
\end{figure}

An example is given in Fig.\ \ref{fig:non-pt},
in which the superpotential, the potential $V_2$ and the wave function obtained
with arbitrary choices for $\xi$ in the intervals to the left and right of the
delta functions are shown for the case $\gamma = 0.3$ and the removal of the
ground state of $\mathcal{H}_1$. One clearly recognises that neither $V_2$ nor
the wave function $\phi_0^{(2)}$ are $\mathcal{PT}$ symmetric. Nevertheless, the
spectrum obtained in this way is identical to that shown in Fig.\
\ref{fig:energies_removed_phi_1}.

\section{Extension to systems with a weak Gross-Pitaevskii
  nonlinearity}
\label{sec:nonlin}

The Gross-Pitaevskii equation contains a nonlinearity which we did
not consider so far. In the cold and dilute gas forming a Bose-Einstein
condensate the van der Waals interaction can be described correctly by
an s-wave scattering process, and the relevant physical parameter defining
the strength of the interaction is given by the s-wave scattering length $a$.
Since this value can be adjusted close to Feshbach resonances by shifting the
molecular energy levels with an external magnetic field it can be chosen
arbitrarily small. However, it is unlikely that the nonlinearity can be
set exactly to zero. Every experiment will at least be affected by small
perturbations. On the other hand it is also unlikely that the SUSY procedure
for the construction of the Fermionic sector's potential will work in the
nonlinear system since the formalism relies on the linearity of the
Hamiltonian. It is the purpose of this section to show that still good
approximate results can be obtained.

\begin{figure}
  \centering
  \includegraphics[width=0.6\textwidth]{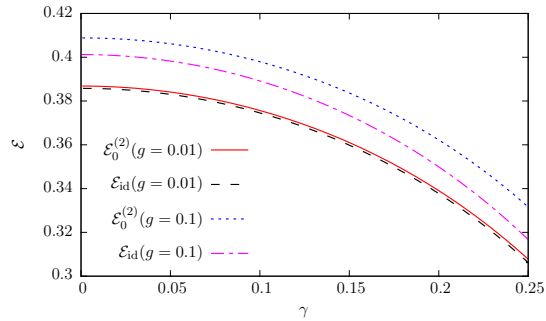}
  \caption{\label{fig:nonlin}Energies $\mathcal{E}_0^{(2)}$ for the nonlinearity
    parameters $g=0.1$ and $g=0.01$ in comparison with the ideal values
    $\mathcal{E}_\mathrm{id}$ according to \eqref{eq:nonlin_ideal}. For both
    values of $g$ the actual energies deviate slightly from the expectation,
    however, the difference is small and the SUSY formalism remains applicable
    with a reasonable quality of
    the results.}
\end{figure}

Including the Gross-Pitaevskii nonlinearity, the potential $V_1$ required in
the SUSY formalism is given by
\begin{equation}
  V_1(x) = \left ( \kappa_n^{(1)} \right )^2 + \nu \delta \left (x-\frac{a}{2}
  \right ) + \nu^\ast \delta \left (x+\frac{a}{2} \right ) + g \left | 
    \phi_n^{(1)} \right |^2 
\end{equation}
with the nonlinearity parameter $g \propto a$. In the units given a small 
nonlinearity means $g \ll 1$. The usage of this $V_1$ in the differential
equation \eqref{eq:deq_superpot} will clearly not lead to a correct partner
system $\mathcal{H}_2$. With this ansatz the nonlinearity will enter into
$V_2$ with the shape of $\phi_n^{(1)}$ and not with that of the solution
$\phi_n^{(2)}$ of the Fermionic sector. Despite this fact the ansatz can be
used for approximate solutions as can be seen in Fig.\ \ref{fig:nonlin}.
In this example the superpotential $\mathcal{W}$ was calculated with the
differential equation \eqref{eq:deq_superpot} for the removal of the ground
state and two different small nonlinearities $g\leq 0.1$. The comparison
between the energy $\mathcal{E}_0^{(2)}$ and the ideal value
\begin{equation}
  \label{eq:nonlin_ideal}
  \mathcal{E}_\mathrm{id} = \mathcal{E}_1^{(1)} - \mathcal{E}_0^{(1)}
\end{equation}
is small but noticeable. 

The method used for constructing the superpotential contains a freedom of
one integration constant as explained in Sect.\ \ref{sec:integration_constant}.
In the nonlinear system this constant is no longer arbitrary. It influences
the energies $\mathcal{E}_n^{(2)}$. We tried to exploit this freedom to
improve the results of the energies or even to enforce the equality
$\mathcal{E}_0^{(2)} = \mathcal{E}_\mathrm{id}$. We found that the values shown
in Fig.\ \ref{fig:nonlin} cannot be improved further. Some choices of the
integration constant even lead to completely asymmetric potentials $V_2$.
However, the important finding of this section is that the nonlinearity
obviously does not completely destroy the concept. In the case of small $g$
the approximation works reasonably well. The energies $\mathcal{E}_0^{(2)}$ are
always slightly above $\mathcal{E}_\mathrm{id}$, but in particular for $g=0.01$
an almost unchanged energy is obtained.

\section{Summary and outlook}
\label{sec:conclusion}

In this paper we studied the supersymmetric extension of the
$\mathcal{PT}$-symmetric double-delta potential. It was possible to show that
the SUSY formalism from non-relativistic quantum mechanics can be used to
remove any of the two states of the original system in an adequately chosen
supersymmetric partner potential. The second state is present in the new
system with exactly the same energy as in the original system but a different
wave function. It is always a symmetric ground state since the original
double-delta potential exhibits only two bound solutions. The application of
the formalism to both states is possible because all solutions of the
non-Hermitian $\mathcal{PT}$-symmetric system ($\gamma \neq 0$) are nodeless.
In the Hermitian double-delta potential the excited state exhibits a node at
the origin and cannot be removed from the spectrum. The corresponding potential
$V_2$ diverges, which could be shown by reducing the non-Hermiticity in our
model potential.

Even exactly at the exceptional point, where only one state is present in
the system, the formalism can be applied. It leads to one wave function in
the partner system with the correct energy. In principle, infinitely many
superpotentials and hence also potentials of the Fermionic sector can be
found. This freedom can be used to either find partner potentials which
preserve the relations of $\mathcal{PT}$ symmetry or lead to cases in which
a non-Hermitian non-$\mathcal{PT}$-symmetric potential exhibits real
eigenvalues or even purely real spectra.

In an extension we investigated whether the formalism can also be used
for the nonlinear Gross-Pitaevskii equation. We found that for small
nonlinearities a partner potential can be constructed, of which the remaining
state's energy is almost unchanged in comparison with its counterpart in the
original system, whereas the other has vanished. Thus, the results are very
similar to the linear case and the most important feature of the supersymmetry
concept is preserved in the nonlinear system. One state is removed from the
spectrum without disturbing the others too much. Unfortunately this is not true
for stronger nonlinearities. Here the simple construction of a superpotential
with a procedure adapted from linear quantum mechanics will fail. However, the
case of strong nonlinearities is the most interesting for the application of
the formalism. It is desirable to remove the $\mathcal{PT}$-broken states
branching off from one of the $\mathcal{PT}$-symmetric eigenstates and
introducing a dynamical instability.

Certainly a way to extend the formalism to arbitrary strengths of the
nonlinearity is the greatest challenge for future work. For this
purpose it could be useful to investigate the many-particle description of
the condensate in second quantisation. It could be promising to try to
factorise the Hamiltonian such that generalised creation and annihilation
operators suitable for the supersymmetry concept can be introduced. It will
be interesting to see whether then two supersymmetric partner systems can
be created and how their mean-field limits are related to each other.


\end{document}